\documentclass{jltp}

\usepackage{graphicx} 
\newcommand{\kk}{{\mathbf k}}

\newcommand{\xx}{{\mathbf x}}

\newcommand{\Psih}{{\hat \Psi}}
\newcommand{\Psihd}{{\hat \Psi}^\dagger}
\newcommand{\eqname}[1]{\label{eq:#1}}
\newcommand{\eq}[1]{(\ref{eq:#1})}

\title{Many-body physics of a quantum fluid of exciton-polaritons in
  a semiconductor microcavity}

\author{Iacopo Carusotto$^{1}$, Michiel Wouters$^{1,2}$, and Cristiano
  Ciuti$^3$}

\address{$^1$ BEC-CNR-INFM and Universit\`a di Trento, I-38050 Povo, Italy \\
$^2$ TFVS, Universiteit Antwerpen, 2610 Antwerpen, Belgium \\
$^3$ Laboratoire Pierre Aigrain, Ecole Normale Sup\'erieure, \\ 
24, rue Lhomond, 75005 Paris, France}

\runninghead{I.Carusotto, M. Wouters, and C. Ciuti}{The quantum fluid
  of polaritons in semiconductor microcavities}

\begin{document}

\maketitle

\begin{abstract}
Some recent results concerning nonlinear optics in semiconductor
microcavities are reviewed from the point of view of the many-body
physics of an interacting photon gas. 
Analogies with systems of cold atoms at thermal
equilibrium are drawn, and the peculiar behaviours due to the
non-equilibrium regime pointed out.
The richness of the predicted behaviours shows the potentialities of
optical systems for the study of the physics of quantum fluids.

\end{abstract}

\section{INTRODUCTION}

So far, the domain of nonlinear optics has remained quite distinct
from the one of quantum fluids, and the concept that
light propagating in a nonlinear optical medium
consists of a gas of photons interacting via the medium nonlinearity
has not been fully exploited yet.
Nonlinear optical effects are generally observed using laser beams
containing a huge number of coherent photons as light sources:
the photon gas is therefore Bose condensed in a single,
macroscopically occupied, mode.
Maxwell equations of classical electromagnetism supplemented by a
nonlinear polarization term are in fact the optical
counterpart of the Gross-Pitaevskii equation for the {\bf C}-number
matter field of an atomic condensate. 
Diluteness of the photon gas is ensured by the small value of the
nonlinearity of most optical media.

In the present paper, we give a brief review of some among the most
recent progress in the nonlinear optical properties of a semiconductor
microcavity system\cite{microcavity_review} which
appears as most suited for studying the many-body behaviour of a
quantum fluid of light. 
Our attention will be focussed on aspects of nonlinear optics that
have a counterpart in quantum fluids, namely
superfluidity\cite{Superfl}, spontaneous symmetry breaking, and the
Goldstone mode~\cite{Goldstone,l_c}. 
In particular, we shall see how the non-equilibrium nature of the
photon gas makes its behaviour much richer than the one of equilibrium
Bose systems.

\section{THE PHYSICAL SYSTEM}

\begin{figure}

\centerline{
\includegraphics[height=3cm,clip]{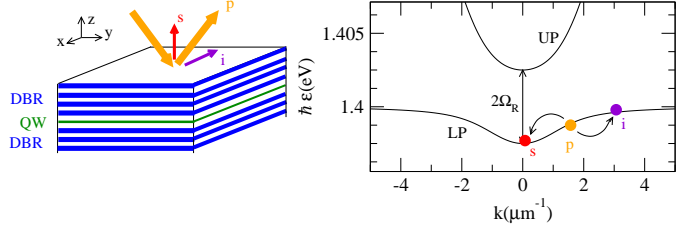}
}

\caption{Left panel: sketch of the microcavity structure and of the
  beams involved in the OPO process. 
Right panel: linear regime polaritonic dispersion and scheme of the OPO
  process.}  
\label{fig:system}
\end{figure}

The physical system we consider\cite{microcavity_review} is sketched in
Fig.\ref{fig:system}a.
Light is confined in the cavity plane by a pair of planar Distributed
Bragg Reflectors (DBR), formed by a stack of alternating $\lambda/4$ layer
of different materials, e.g. AlAs and GaAs.
The active material consists of a thin layer of a smaller gap material,
e.g. InGaAs, present in the cavity layer:
the Coulomb interaction between carriers results in a sharp optical
line corresponding 
to the creation of a hydrogen-like electron-hole pair called exciton.
If the excitonic transition is close to resonance to the cavity mode
and the exciton-photon coupling $\Omega_R$ exceeds the excitonic and
photonic linewidths $\gamma_{X,C}$, the eigenmodes of the system are
{\em exciton-polaritons}, i.e. linear superpositions of cavity photon
and quantum well exciton. 
This regime is conventionally called the {\em strong coupling}
regime.
In the low-density limit, polaritons satisfy Bose statistics.
A simple description of the system is provided by the following
second quantized Hamiltonian~\cite{Superfl,OPO_th}: 
\begin{eqnarray}
  H&=&\!\!\!\!\int\!d\xx\,\!\!\!\!\sum_{ij=\{X,C\}}\!\!\!
\Psihd_{i}(\xx) \,{\mathbf
  h}^{0}_{ij}\,\Psih_{j}(\xx)
+\frac{\hbar g}{2}\int\! d\xx\,\Psihd_{X}(\xx)\,\Psihd_{X}(\xx)\,
\Psih_{X}(\xx)\,\Psih_{X}(\xx)+ \nonumber\\
&+&\!\!\!\!\int\!d\xx\,\hbar \left(F_{p}(\xx,t) \,\Psihd_{C}(\xx)+
F^*_{p}(\xx,t) \,\Psih_{C}(\xx) \right)+H_{\rm bath}.
\eqname{Hamilt}
\end{eqnarray}
The (bosonic) quantum field operators $\Psi_{C,X}(\xx)$ describing the
cavity-photon $C$ and the exciton $X$ fields are defined in the 2D
cavity plane.
The single-particle Hamiltonian ${\mathbf h}^0$
\begin{equation}
  \label{eq:h_0}
  {\mathbf h}^0=
\hbar \left(
\begin{array}{cc}
\omega_{X}(-i\nabla) + V_X(\xx) & \Omega_R \\
\Omega_R & \omega_C(-i\nabla)+V_C(\xx)
\end{array}
\right),
\end{equation}
involves the cavity-photon and exciton dispersions
$\omega_{C,X}(\kk)$, and the electric-dipole exciton-photon coupling
$\Omega_R$. 
In the absence of confining potential $V_{C,X}(\xx)=0$, the eigenmodes of
${\mathbf h}^0$ correspond to the two polariton 
branches, the lower (LP) and the upper (UP) polariton. Their
dispersion is plotted in Fig.\ref{fig:system}b.
The excitonic content of the polariton is responsible for their binary
interactions, which provide the optical nonlinearity of 
the system.
The possibility of applying a spatial confinement potential $V_{C,X}(\xx)$ to
polaritons by suitably nanostructuring the cavity has been recently
demonstrated e.g. in Ref.\onlinecite{plots}, which opens the way to polariton optics
applications\cite{blockade}.

All the terms in \eq{Hamilt} mentioned so far have analogs in the
second-quantized formulation of ultra-cold atom system.
New physics arises from the last two terms, which respectively
describe injection of polaritons into the cavity by a coherent light
beam incident onto the front mirror, and their dissipation into the
bath of other modes: $F_p(x,t)$ is proportional to the incident field
amplitude. 
The main decay channel for the photon mode corresponds to the emission 
of radiation through the non-perfectly reflecting mirrors, while many
different decay channels are available for the exciton mode, 
e.g. scattering on phonons, non-radiative exciton recombination at
defects, excitation induced decoherence, etc.

\section{HOW TO CREATE AND OBSERVE A POLARITON GAS}

Creation, control and diagnostics of the polariton gas in the
microcavity are generally performed by optical means through the
cavity mirrors.
This provides a clear advantage over most other quantum fluids, 
as optical injection of polaritons combines experimental simplicity
with a wide flexibility.
A variety of flow configurations can be generated by driving
the cavity with a laser field with the suitable spatial and temporal
profile $F_p(\xx,t)$. 
Homogeneous and stationary flow along the cavity plane are obtained
using a tilted pump beam of in-plane wavevector $\kk_p$. 
More complex configurations showing e.g. vortical flow patterns are
obtained using Laguerre-Gauss shaped pump beams\cite{optical_vortices}.
It is interesting to stress the fact that polaritons are automatically
created in the cavity in a coherent state; this state satisfying the
Penrose-Onsager criterion\cite{book,Huang}, it can be seen as a
(non-equilibrium) Bose condensate of polaritons. 

The diagnostics of the polariton gas can be performed by optical
means too, by measuring the light emitted by the cavity. 
Once direct reflection of the pump laser is eliminated, the
(operator-valued) electric field amplitude of the emitted light from
the cavity is in fact proportional to the photonic component of the
quantum polariton field~\cite{W-QMC}.
All correlation functions of the in-cavity polariton field are
therefore directly translated into the corresponding ones of the
extra-cavity emitted radiation, which are easily measured by means of
standard optical interferometric techniques\cite{Baas_PRL}. 

The first example of many-body property of the polariton gas to be
studied has been superfluidity.
A generalized Gross-Pitaevskii equation has been
written~\cite{Superfl}:
{\small
\begin{equation}
  \label{eq:GPE}
  i\,\frac{d}{dt}
\left(
  \begin{array}{c}
\psi_{X}\\ \psi_{C}
  \end{array}
\right)= \left[ {\mathbf h}^0 + \left(
\begin{array}{cc}
g|\psi_X|^2 - i \gamma_X& 0 \\ 0 & -i
\gamma_C
  \end{array}
\right) \right] \left(
  \begin{array}{c}
\psi_{X} \\ \psi_{C}
  \end{array}
\right)
+\left(
  \begin{array}{c}
0 \\ F_{p} \,
  \end{array}
\right)
\end{equation}
}
and used to derive a non-equilibrium formulation\cite{Superfl} of the
Landau criterion~\cite{book,Huang} for superfluidity as a function of
the pump beam intensity, frequency, and wavevector.
When a very dilute polariton gas flowing along the cavity plane
scatters on the sample imperfections, the typical ring-shaped
profile of resonant Rayleigh scattering appears in the far-field
emission. 
On the other hand, when the polariton gas becomes denser and the sound
speed in it overcomes the flow velocity, the defect is no longer able
to create excitations in the polariton fluid and the ring-shaped feature
disappears from the far-field emission.


\section{PARAMETRIC THRESHOLD AS A POLARITON BEC}
Imposing the coherence from the outside is not the only way of
obtaining a coherent polariton gas: several recent experiments have
shown the spontaneous appearance of coherence under a sufficiently
strong pumping\cite{OPO,BEC_exp}.
This {\em spontaneous coherence} is not imprinted from the outside,
and often even appears in modes different from 
the ones which are optically pumped. 
In this respect, the system mostly ressembles a laser above threshold,
with the important difference that a continuum of modes available to
the emission, so that the field is not frozen in a cavity mode, but
has non-trivial spatial dynamics along the cavity plane. 
Although we will refer here to the specific example
(Fig.\ref{fig:system}) of parametric
oscil\-la\-tion\cite{OPO_th,OPO}, for which a complete and microscopic
theoretical 
description is available\cite{W-QMC}, our conceptual framework can be
extended to any other scheme showing spontaneous
coherence\cite{BEC_exp,BEC_th}.
Pump polaritons are injected at
$\kk_p$ and then converted into a pair of signal and idler ones at
$\kk_{s,i}\neq \kk_p$ by the excitonic nonlinearity.
Above a certain threshold, the emission at $\kk_{s,i}$
goes from thermal to coherent as a consequence of the spontaneous
breaking of a $U(1)$ 
signal/idler phase symmetry: the phase of the
signal/idler emissions become coherent over long distances.
According to the Penrose-Onsager criterion of Bose
condensation\cite{Huang}, one 
can therefore speak of a non-equilibrium Bose condensate.

This has been observed in Wigner-Quantum Monte Carlo simulations of
the polariton field\cite{W-QMC}.
The behaviour of the first- and second-order coherence
functions of the signal emission results in the threshold region
qualitatively identical to the one of a Bose gas at equilibrium in the
vicinity of the Bose-Einstein condensation point.
While approaching the critical point from below, the first-order
coherence length increases, becomes macroscopic, and finally diverges
at criticality. 
Above threshold, coherence extends across the whole sample. 
On the other hand, the second-order coherence function below threshold
shows the typical Hanbury-Brown and Twiss (HBT) bosonic bunching upto   
distances of the order of the coherence length, and then goes back to 
$1$. Above threshold, it is instead flat and equal to $1$.

First experimental observations of these facts have been recently
re\-ported\cite{Baas_PRL}: the long-distance coherence of the  
signal emission is inferred from the interference fringes that are
obtained when light extracted at two distant points is made to beat.
The monomode nature of the intensity noise of the signal emission
above the threshold has been also established, which has ruled out the
occurrence of effects of HBT type. 

\section{NEW FEATURES DUE TO NON-EQUILIBRIUM}

Despite the many similarities, fundamental differences can arise
because of the non-equilibrium nature of the system under investigation. 
Standard techniques of equilibrium statistical mechanics such as the
Boltzmann law $p\propto \exp(-E/k_B T)$ can not be applied,
and the stationary state of the polariton fluid is determined by a
dynamical balance between driving (i.e. injection) and dissipation
(i.e. losses). 
From this respect, the long-distance spatial coherence of the signal
emission is an optical analog of the regular periodic 
arrangement of B\'enard cells in heat convection\cite{pattern}.  

The {\em Goldstone mode} due to the spontaneously broken
  $U(1)$ symmetry is not a propagating sound-like mode, but rather a
  diffusive one\cite{Goldstone}. This effect can be experimentally
  observed by probing the elementary excitations around the
  parametrically  oscillating state with an additional laser beam.
 Pioneering luminescence experiments in this direction have been reported\cite{Goldstone_exp}. 

At equilibrium, free-energy minimization arguments force
  the BEC to form in the lowest-energy state.
  Far from equilibrium, this constraint is lifted, and techniques
  mutuated from pattern formation can be used\cite{pattern_our} to
  determine the shape of the condensate 
  mode as a function of the geometry and the pump frequency.

The spatial decay of coherence in non-equilibrium 1D
  systems\cite{dasbach} is 
  exponential even in the absence of thermal effects, with a coherence
  length $\ell_c$ inversely proportional to the damping rate $\gamma$ and 
  to the second derivative of the {\em imaginary} part of the
  {\em Goldstone mode} dispersion. This, in contrast to the equilibrium case
  where $\ell_c$ is inversely proportional to the temperature $T$ and
  to the second derivative of the {\em real} part of the {\em free
  boson} dispersion\cite{l_c}.

\section*{ACKNOWLEDGMENTS}

M.W. acknowledges financial support from the FWO-Vlaanderen in the form
of a ``mandaat  Postdoctoraal Onderzoeker''.
LPA-ENS is a "Unit\'{e} Mixte de Recherche Associ\'{e} au CNRS (UMR
8551) et aux Universit\'{e}s Paris 6 et 7"


\begin{thebibliography}{99}


\bibitem{microcavity_review} B. Deveaud (Ed.), {\em Physics of
  semiconductor microcavities}, Special issue of: {\em
  Phys. Stat. Sol. B} {\bf 242},  2145-2356 (2005). 

\bibitem{Superfl} I. Carusotto and C. Ciuti, {\em Phys. Rev. Lett.}
  {\bf 93}, 166401 (2004)

\bibitem{Goldstone} M. Wouters and I. Carusotto, preprint
  {\em cond-mat/0606755}

\bibitem{l_c} M. Wouters and I. Carusotto, preprint {\em
  cond-mat/0512464}

\bibitem{OPO_th} C. Ciuti, P. Schwendimann, and A. Quattropani,  {\em
  Semicond. Sci. Technol.} {\bf 18}, S279-S293 (2003) and references therein.

\bibitem{plots} O. El Da\"if {\em et al.}, {\em Appl. Phys. Lett.}
  {\bf  88}, 061105 (2006)

\bibitem{blockade} A. Verger {\em et al.}, {\em Phys. Rev. B} {\bf
  73}, 193306  (2006)


\bibitem{optical_vortices} D. Rozas, Z. S. Sacks, and
  G. A. Swartlander, {\em Phys.Rev.Lett.} {\bf 79}, 3399 (1997).


\bibitem{book} L.P. Pitaevskii and S. Stringari, {\sl Bose-Einstein
  Condensation}, Clarendon Press, Oxford (2003).


\bibitem{Huang} K. Huang, {\em Statistical Mechanics}, Wiley, New
  York, 1997. 



\bibitem{W-QMC} I. Carusotto and C. Ciuti, {\em Phys. Rev. B} {\bf
  72}, 125335 (2005) 

\bibitem{Baas_PRL} A. Baas, {\em et al.}, {\em Phys. Rev. Lett.} {\bf
  96}, 176401 (2006) 

\bibitem{OPO} R. M. Stevenson {\sl et al.}, {\em Phys. Rev. Lett.} {\bf
  85}, 3680 (2000); R. Houdr\'e {\sl et al.}, {\em Phys. Rev. Lett.}
  {\bf 85}, 2793 (2000).



\bibitem{BEC_exp} M. Richard, {\em et al.}, {\em Phys. Rev. B} 72,
  201301 (2005);  H. Deng, {\em et al.}, {\em Science} {\bf 298}, 199
  (2002); J. Kasprzak {\em et al,}, {\em Nature} {\bf 443}, 409 (2006) 

\bibitem{BEC_th}  D. Porras, {\em et al.}, {\em Phys. Rev. B} 66,
  85304 (2002); F. P. Laussy, {\em et al.}, {\em  Phys. Rev. Lett.}
  {\bf 93}, 016402 (2004); D. Sarchi and V. Savona, preprint {\em
  cond-mat/0603106}; F. M. Marchetti, {\em et al.}, {\em
  Phys. Rev. Lett.} {\bf 96}, 066405 (2006).
\bibitem{pattern} M. C. Cross and P. C. Hohenberg, {\em
  Rev. Mod. Phys.} {\bf 65}, 851 (1993) 

\bibitem{Goldstone_exp} P. G. Savvidis, {\em et al.}, {\em
  Phys. Rev. B}, {\bf 64}, 075311 (2001)
\bibitem{pattern_our} M. Wouters and I. Carusotto, in preparation.

\bibitem{dasbach} G. Dasbach, {\em et al.}, {\em Phys. Rev. B} {\bf
  71}, 161308(R) (2005). 




\end{thebibliography}
\end{document}